\documentclass[reqno,11pt]{amsproc}
\usepackage{footnpag}
\usepackage{amsthm}
\usepackage{amsmath}
\usepackage{amsfonts}
\usepackage{amssymb}
\usepackage{amsmath,amssymb}
\usepackage{graphicx}

\newtheorem{theorem}{Theorem}
\numberwithin{equation}{section}
\numberwithin{theorem}{section}

\newtheorem{definition}[theorem]{Definition}

\newtheorem{lemma}[theorem]{Lemma}

\newtheorem{proposition}[theorem]{Proposition}

\newtheorem{remark}[theorem]{Remark}

\newcommand{\FORALL} {{\hbox{$\hskip 11mm \forall \;$}}}
\newcommand{\Ascr} {{\mathcal A}}
\newcommand{\Cscr} {{\mathcal C}}
\newcommand{\Dscr} {{\mathcal D}}
\newcommand{\dsp}{\displaystyle}

\begin{document}

\title{Reconstructing the potential for the 1D Schr\"odinger equation from boundary measurements}


\author{ S. A. Avdonin}
\address{S. A. Avdonin, Department of Mathematics and Statistics University of Alaska
Fairbanks, AK 99775-6660, USA} \email{saavdonin@alaska.edu}
\author{ V. S. Mikhaylov}
\address{V. S. Mikhaylov, St.Petersburg   Department   of   V.A.Steklov    Institute   of   Mathematics
of   the   Russian   Academy   of   Sciences, 7, Fontanka, 191023
St.Petersburg, Russia} \email{vsmikhaylov@pdmi.ras.ru}
\author{K. Ramdani}
\address{K. Ramdani, INRIA Nancy Grand-Est (CORIDA)
615 rue du Jardin Botanique 54600, Villers-les-Nancy, France}
\email{karim.ramdani@inria.fr}

\maketitle
\date{November 09, 2011}


\begin{center}
    {\bf Abstract.} We consider the inverse problem of the determining the potential in the dynamical
    Schr\"odinger equation on the interval by
    the measurement on the whole boundary. Provided that source is \emph{generic} using the Boundary
    Control method we recover the spectrum of the problem from the
    observation at either left or right end points.
    Using the specificity of the one-dimensional situation we
    recover the spectral function, reducing the problem to the
    classical one which could be treated by known methods. We
    adapt the algorithm to the situation when only the finite
    number of eigenvalues are known and provide the result on
    the convergence of the method.
\end{center}

\section{Introduction}
We consider the problem of determining the potential for a one dimensional Schr\"odinger equation from two boundary measurements. More precisely, given  $q\in L^1(0,1)$ (without loss of generality, we assume that $q$ is real-valued) and $a\in H^1_0(0,1)$, we consider the following initial boundary value problem:
\begin{equation}
    \label{eq_1}
    \left\{
    \begin{array}{ll}
        \dsp iu_t(x,t)-u_{xx}(x,t)+q(x)u(x,t)=0 &\qquad t>0,\quad 0<x<1\\
        \dsp u(0,t)=u(1,t)=0&\qquad t>0,\\
        \dsp u(x,0)=a(x) & \qquad 0<x<1.
    \end{array}
    \right.
\end{equation}
Assuming that the initial data $a$ is unknown, the inverse problem we are interested in is to determine the potential $q$ from the trace of the derivative of the solution $u$ to
(\ref{eq_1}) on the boundary:
$$\{r_0(t),r_1(t)\}:=\{u_x(0,t),u_x(1,t)\}, \FORALL t\in (0,T),$$
where $T>0$ is a fixed constant. Once the potential has been determined, we can use the method of iterative observers recently proposed in \cite{KTW} to recover the initial data.

The multidimensional inverse problems for the Schr\"odinger
equation by one measurement were considered in
\cite{BP,BP1,BC,MOR}. Using techniques based on Carleman estimates, the authors established global uniqueness and stability results for different geometrical conditions on the domain. We also point out the approach developed by Boumenir and Tuan for the heat equation in \cite{BT3,BT, BT2}. Using observation at the one end of the interval for $t\in (0,\infty)$, the authors were able to recover the spectrum of the string. Next, choosing another boundary condition, they solved the inverse problem
from the two recovered spectra by the classical Levitan-Gasymov method
\cite{L,LG}.

In this paper we offer a different approach. More precisely, let us introduce
the unbounded operator $\Ascr$ on $L^2(0,1)$ defined by
\begin{equation}\label{eqA}
\Ascr\phi := -\phi'' +q\phi,\FORALL \phi\in \Dscr(\Ascr):=H^2(0,1)\cap H^1_0(0,1).
\end{equation}
Using the Boundary Control Method (denoted BCM in the rest of the paper, see \cite{AGM,B07}), we first show that the eigenvalues of $\Ascr$
can be recovered from the data $r_0$ (or $r_1$). To achieve this, we derive a generalized eigenvalue problem involving an integral operator (see \eqref{Int_eq}), whose resolution leads to the recovery of the eigenvalues of $\Ascr$.
Then, using the peculiarity of the one dimensional
case, we recover the spectral function associated with $\Ascr$, reducing the initial inverse problem to the more ``classical" one of recovering an unknown potential from spectral data. We can then use either Gelfand-Levitan, Krein or Boundary Control
Methods (see \cite{AM}).
\begin{remark}
The method of solving the inverse problem presented in this paper
could be applied to the case of wave and parabolic equations on
the interval as well.
\end{remark}
The paper is organized as follows : in Section \ref{sect_BCM}, we detail the different steps of our algorithm.
In particular, we describe how to recover the spectral data of $\Ascr$ from the boundary data and point out two
methods that can be used to recover the potential. In Section \ref{sect_stability}, we show how to adapt our
algorithm to the more realistic situation where not all but only a finite number of eigenvalues of $\Ascr$ are known.

\section{Inverse problem, application of the BCM}
\label{sect_BCM}

\subsection{From boundary data to spectral data}
It is well known that the selfadjoint operator $\Ascr$ defined by \eqref{eqA} admits a family of eigenfunctions
$\{\phi_k\}_{k=1}^\infty$, associated to a sequence of eigenvalues $\lambda_k\rightarrow +\infty$, forming a orthonormal basis of $L^2(0,1)$. Using the method of
moments, we can represent the solution of (\ref{eq_1}) in the form
\begin{equation}
\label{Repr} u(x,t)=\sum_1^\infty a_ke^{i\lambda_k t}\phi_k(x),
\end{equation}
where $a_k$ are (unknown) Fourier coefficients:
\begin{equation}
a_k=\left(a,\phi_k\right)_{L^2(0,1;dx)}.
\end{equation}
\begin{definition}
We call the initial data $a\in H^1_0(0,1)$ generic if $a_k\not=0$ for all $k\ge 1$.
\end{definition}

For the boundary data $r_0$, $r_1$, we have the representation
\begin{equation}
\left\{r_0(t),r_1(t)\right\}=\left\{\sum_{k=1}^\infty
a_ke^{i\lambda_k t}\phi_k'(0),\sum_{k=1}^\infty a_ke^{i\lambda_k
t}\phi_k'(1)\right\}.\label{RT}
\end{equation}
In particular, one can note that $r_0, r_1\in L^2(0,T)$.
Throughout the paper, we always assume that the source $a$ is generic.
The reason for this requirement can be observed from (\ref{RT}):
if we assume that $a_n=0$ for some $n$, then from the
representation (\ref{RT}) it is easy to see that the response
function does not contain any information about the triplet
$\{\lambda_n,\phi_n'(0),\phi_n'(1)\}$, and, consequently the
potential which would be possible to recover from $\left\{r_0(t),r_1(t)\right\}$ would not be unique.

Using the method from \cite{AGM} we can recover the spectrum
$\lambda_k$ and products $a_k\phi_k'(0)$ and $a_k\phi_k'(1)$ by
the following procedure. We construct the operator $C_0^T:
L^2(0,T)\longrightarrow L^2(0,T)$ by the rule:
\begin{equation}
\label{C_T_0} (C_0^T)f(t)=\int_0^T
r_0(2T-t-\tau)f(\tau)\,d\tau,\qquad 0\leqslant t\leqslant T.
\end{equation}
and consider the following generalized eigenvalue problem : Find $(\mu,f)\in {\mathbb C}\times L^2(0,T)$, $f\neq 0$, such that
\begin{equation}
\label{Int_eq} \int_0^T \dot r_0(2T-t-\tau) f(\tau) = \mu
\int_0^T r_0(2T-t-\tau)f(\tau)\,d\tau,\qquad
0\leqslant t\leqslant T.
\end{equation}
Then, one can prove that the above problem admits a countable set of
solutions $(\mu_n,f_n)$, $n\geqslant 1$. Moreover, the eigenvalues $\mu_n$ are precisely given by the values $i\lambda_n$, where $\lambda_n$ are
the eigenvalues of $\Ascr$ and the family
$\{f_n(t)\}_{n=1}^\infty$ is biorthogonal to
$\{e^{i\lambda_k(T-t)}\}_{n=1}^\infty$.

Next, we consider the equation
of the form (\ref{Int_eq}) with $r_0(t)$ replaced by its complex
conjugate $\overline{r_0(t)}$. This equation yields the sequence $\{
-i \lambda_k,g_k(t)\}_{k=1}^\infty$,
($-i\lambda_n=\overline{\mu}_n$). Let us normalize functions
$f_k$, $g_k$ by the rule:
\begin{equation}
\label{Normal}
\delta_{nk}=\left(C_0^Tf_n,g_k\right)=\int_0^T\int_0^T
r_0(2T-t-\tau)f_n(\tau)\overline{g_k(t)}\,d\tau\,dt
\end{equation}
and introduce constants $\gamma_k$, $\beta_k$ defined by:
\begin{eqnarray}
\gamma_k=\int_0^T r_0(T-\tau)f_k(\tau)\,d\tau,  \label{gamma_k}\\
\beta_k=\int_0^T
r_0(T-\tau)\overline{g_k(\tau)}\,d\tau.\label{beta_k}
\end{eqnarray}
Then the product $a_k\phi_x'(0)$ could be evaluated by
\begin{equation}
\label{coeff_left} a_k\phi_k'(0)=\gamma_k\beta_k.
\end{equation}
Similarly we can introduce the integral operator $C_1^T$
associated with the response at the right endpoint $r_1(t)$, and
repeat the procedure described above to find quantities
$a_k\phi_k'(1)$.

Summing up, using the method from \cite{AGM} we are able to recover the
eigenvalues $\lambda_k$ of $\Ascr$ and the products $\phi_k'(0)a_k$ and
$\phi_k'(1)a_k$. As a result, we can say that we recovered the
spectral data consisting of
\begin{equation}
\label{Data}
\Dscr:=\left\{\lambda_k,\frac{\phi_k'(1)}{\phi_k'(0)}\right\}_{k=1}^\infty.
\end{equation}
Precisely this data was used in \cite{PT}, where the authors shown
the uniqueness for the inverse spectral problem and provided the
method of recovering the potential. Instead of doing this, we
recover the spectral function associated to $\Ascr$ and thus reduce the inverse source problem to the classical one of determining an unknown potential from spectral data.

Given $\lambda\in {\mathbb C}$, we introduce the solution $y(\cdot,\lambda)$ of the following Cauchy problem on $(0,1)$:
\begin{equation}
\label{Aylambday} - y''(x,\lambda) + q(x)y(x,\lambda)=\lambda y(x,\lambda),\qquad 0<x<1,
\end{equation}
\begin{equation}
\label{Cauchy_cond} y(0,\lambda)=0,\quad y'(0,\lambda)=1.
\end{equation}
Then the eigenvalues of the Dirichlet problem  of $\Ascr$ are exactly the zeroes of the function $y(1,\lambda)$, while a family of normalized corresponding eigenfunctions is given by
$\phi_k(x)=\dfrac{y(x,\lambda_k)}{\|y(\cdot,\lambda_k)\|}$. Thus we
can rewrite the second components in $\Dscr$ in the following way:
\begin{equation}
\label{A_k}
\frac{\phi_k'(1)}{\phi_k'(0)}=\frac{y'(1,\lambda_k)}{y'(0,\lambda_k)}=y'(1,\lambda_k)=:A_k.
\end{equation}
Let us denote by dot the derivative with respect to $\lambda$. We
use the following fact (see \cite{PT}), which is also valid for
$q\in L^1(0,1)$:
\begin{lemma}\label{lem1}
If $\lambda_n$ is an eigenvalue of $\Ascr$, then
\begin{equation}
\|y(\cdot,\lambda_n)\|^2_{L^2}=y'(1,\lambda_n)\dot y(1,\lambda_n)
\end{equation}
\end{lemma}
The lemma below which can be found in \cite{PT} for the case
$q\in L^2(0,1)$ holds true for $q\in L^1(0,1)$ as well. The
meaning of it is that the function $y(1,\lambda)$ is completely
determined by its zeroes, which are precisely the eigenvalues of $\Ascr$.
\begin{lemma}\label{lem2}
For $q\in L^1(0,1)$ the following representations
hold
$$
y(1,\lambda)=\prod_{k\geqslant 1}\frac{\lambda_k-\lambda}{k^2\pi^2}
$$
$$
\dot y(1,\lambda_n)=-\frac{1}{n^2\pi^2}\prod_{k\geqslant 1,
k\not=n}\frac{\lambda_k-\lambda_n}{k^2\pi^2}.
$$
\end{lemma}
Lemma \ref{lem1} and \ref{lem2} imply that the data $\Dscr$ we have recovered (see \eqref{Data}) allow us to evaluate the norm
\begin{equation}
\label{alpha_k} \|y(\cdot,\lambda_n)\|^2_{L^2}=A_nB_n=:\alpha_n^2,
\end{equation}
where we denoted
\begin{equation}
\label{B_k} B_n:=-\frac{1}{n^2\pi^2}\prod_{k\geqslant 1,
k\not=n}\frac{\lambda_k-\lambda_n}{k^2\pi^2},
\end{equation}

\subsection{Reconstructing the potential from spectral data}
The set of pairs \newline
$\{\lambda_k,\|y(\cdot,\lambda_k)\|^2_{L^2}\}_{k=1}^\infty$ is a
``classical'' spectral data. The potential can thus be recovered
by Gelfand-Levitan, Krein method or the BCM (see \cite{AM}). Below
we outline two possible methods of recovering the potential.

We introduce the spectral function associated with $\Ascr$:
\begin{equation}
\label{sp_func} \rho(\lambda)=\left\{\begin{array}l
-\sum\limits_{\lambda\leqslant\lambda_k\leqslant
0}\frac{1}{\alpha_k^2}\quad \lambda\leqslant 0, \\
\sum\limits_{0<\lambda_k\leqslant\lambda}\frac{1}{\alpha_k^2}\quad
\lambda
> 0,
\end{array}\right.
\end{equation}
which is a monotone increasing function having jumps at the points
of the Dirichlet spectra. The regularized spectral function is
introduced by
\begin{equation}
\label{reg_sp_func}
\sigma(\lambda)=\left\{\begin{array}l \rho(\lambda)-\rho_0(\lambda)\quad \lambda\geqslant 0, \\
\rho(\lambda)\quad \lambda < 0,
\end{array}\right. \quad
\rho_0(\lambda)=\sum\limits_{0<\lambda_k^0\leqslant\lambda}\frac{1}{(\alpha_k^0)^2}\quad
\lambda
> 0,
\end{equation}
where $\rho_0$ is the spectral function associated with the operator $\Ascr$ when $q\equiv 0$. In the definition above eigenvalues and norming coefficients are
$\lambda_k^0=\pi^2 k^2$, $(\alpha_k^0)^2=\frac{1}{2\pi^2k^2}$ and
the solution to \eqref{Aylambday}-\eqref{Cauchy_cond} for $q\equiv 0$ is
$y_0(x,\lambda)=\dfrac{\sin{\sqrt{\lambda}x}}{\sqrt{\lambda}}$.

Let us fix $\tau\in (0,1]$ and  introduce the kernel $c^\tau(t,s)$
by the rule (see also \cite{AM}):
\begin{equation}
\label{C_T_eqiv} c^\tau(t,s)=\int_{-\infty}^\infty
\frac{\sin{\sqrt{\lambda}(\tau-t)}\sin{\sqrt{\lambda}(\tau-s)}}{\lambda}\,d\sigma(\lambda),\quad
s,t\in [0,\tau],
\end{equation}
Then so-called connecting operator (see \cite{B07,AM}) $C^\tau:
L^2(0,\tau)\mapsto L^2(0,\tau)$ is defined by the formula
\begin{equation}
\label{r-c2} ({C}^\tau f)(t)=(I+\Cscr^\tau)f(t)=f(t)+\int_0^\tau
c^\tau(t,s)f(s)\,ds\,, \ 0<t<\tau\,,
\end{equation}
Using the BCM leads to the following: for fixed $\tau\in (0,1)$ one solves the equation
\begin{equation}
\label{BC_equation} (C^\tau f^\tau)(t)=\tau-t,\quad 0<t<\tau,
\end{equation}
Setting
\begin{equation}
\label{qmu_equation} \mu(\tau):= f^\tau(+0),
\end{equation}
and then varying $\tau\in(0,1)$, the potential at the point $\tau$ is recovered by
\begin{equation}
\label{mainq} q(\tau)=\frac{\mu''(\tau)}{\mu(\tau)}\,.
\end{equation}

We can also make use of the Gelfand-Levitan theory. According to this approach, for $\tau\equiv 1$, the kernel $c^{\tau}$ satisfies the following
integral equation with unknown $V$:
\begin{equation}
\label{G_LN} V(y,t)+c^{\tau}(y,t)+\int_y^\tau
c^\tau(t,s)V(y,s)\,ds=0,\quad 0<y<t<1.
\end{equation}
Solving the equation (\ref{G_LN}) for all $y \in (0,1)$ we can
recover the potential using
\begin{equation}
\label{Rec_potenc_GL} q(y)=2\frac{d}{dy}V(\tau-y,\tau-y).
\end{equation}

Once the potential has been found, we can recover the eigenfunctions
$\phi_k$, the traces $\phi_k'(0)$ and using (\ref{coeff_left}),
the Fourier coefficients $a_k$, $k=1,\ldots,\infty$. Thus, the
initial state can be recovered via its Fourier series. We can
also use the method of observers (see \cite{KTW}).

\subsection{The algorithm}

\begin{itemize}
\item[1)] Take $r_0(t):=u_x(0,t)$ and solve the generalized
spectral problem (\ref{Int_eq}). Denote the solution by
$\{\mu_n,f_n(t)\}_{n=1}^\infty$ and note the connection with the
spectra of $\Ascr$: $\lambda_n=-i\mu_n$.

\item[2)] Take the function $\overline{ r_0(t)}$ and repeat step
one. This yields the sequence $\{\overline{\mu_n},g_n(t)\}_{n=1}^\infty$.

\item[3)] Define the operator $C^T_0$  by (\ref{C_T_0}) and
normalize $f_n$, $g_n$ according to equation (\ref{Normal}):
$(C^T_0f_n,g_n)=1$.

\item[4)] Find the quantities  $a_n\phi_n'(0)$ by (\ref{gamma_k}),
(\ref{beta_k}) and (\ref{coeff_left}).

\item[5)] Repeat steps 1)--4) for the function
$r_1(t):=u_x(1,t)$ to find $a_k\phi_k'(1)$.

\item[6)] Define the spectral data $\Dscr$ by (\ref{Data}) and find
the norming coefficients $\alpha_k$ by using (\ref{alpha_k}) and
(\ref{A_k}), (\ref{B_k}).

\item[7)] Introduce the spectral function $\rho(\lambda)$,
the regularized spectral function $\sigma(\lambda)$ and the kernel
$c^\tau$ respectively defined by (\ref{sp_func}), (\ref{reg_sp_func}) and
(\ref{C_T_eqiv}).

\item[8)] Solve the inverse problem by either BCM using
(\ref{BC_equation}), \eqref{qmu_equation}, (\ref{mainq}) or Gelfand-Levitan method using equations (\ref{G_LN}), (\ref{Rec_potenc_GL}).

\item[9)] Use the method of iterative observers described in \cite{KTW} or Fourier series to recover the initial data.
\end{itemize}

Our approach yields the following uniqueness result for the
inverse problem for 1-d Schr\"odinger equation:

\begin{theorem}
Let the source $a\in H^1_0(0,1)$ be generic and $T$ be an
arbitrary positive number. Then the potential $q\in L^1(0,1)$ and
the initial data are uniquely determined by the observation
$\{u_x(0,t),u_x(1,t)\}$ for $t\in (0,T)$.
\end{theorem}

The method could be applied to the inverse problem for the wave
and parabolic equations with the potential on the interval as
well. The details of the recovering the spectrum $\lambda_k$ and
the quantities $a_k\phi_k'(0)$, $a_k\phi_k'(1)$ could be found in
\cite{AGM}. The following important remark, connected with the
types of the controllability of the corresponding systems, should
be taken into the consideration:
\begin{remark}
The time $T$ of the observation could be arbitrary small for the
case of Schr\"odinger and parabolic equations and is doubled the
length of an interval ($T=2$ in our case) for the wave equation
with the potential.
\end{remark}
For the details see \cite{AGM}.

\section{Stability of the scheme : the case of truncated spectral data}
\label{sect_stability}
In view of applications, in this section we consider the case where only a finite number of eigenvalues of $\Ascr$ are available. More precisely, let us assume that we recovered the exact values of the first $N$ eigenvalues $\lambda_n$ and traces $A_n$, for $n=1,\ldots,N$. Then we can introduce the approximate normalizing coefficients
$\widetilde\alpha_{n,N}$ by the rule
\begin{equation}
\label{approx_coeff} \widetilde\alpha_{n,N}=A_n\widetilde B_{n,N},\quad
\text{where}\quad\widetilde
B_{n,N}:=-\frac{1}{n^2\pi^2}\prod_{k\geqslant 1,
k\not=n}^N\frac{\lambda_k-\lambda_n}{k^2\pi^2}
\end{equation}
Then we can estimate
\begin{equation}
\label{alpha_est}
|\alpha_n-\widetilde\alpha_{n,N}|\leqslant|A_n||\widetilde
B_{n,N}|\left|1-\prod_{k\geqslant
N+1}^\infty\frac{\lambda_k-\lambda_n}{k^2\pi^2}\right|
\end{equation}
Since the infinite product (\ref{B_k}) converges, the right hand
side of the above inequality is small enough as $N$ is getting
bigger, provided $n$ is fixed. But the following remark should be
taken into the account. Let us remind the following asymptotic
formulas for the eigenvalues and norming coefficients:
\begin{eqnarray}
\lambda_k=\pi^2k^2+c+O\left(\frac{1}{k^2}\right),\quad k\to
\infty,\label{eigen_exp}\\
\alpha_k^2=\frac{1}{2\pi^2 k^2}+O\left(\frac{1}{k^4}\right),\quad
k\to\infty,\label{norm_coeff_exp}
\end{eqnarray}
where $c=\dsp \int_0^1q(s)\,ds$. Then the product in the right hand
side of (\ref{alpha_est}) can be rewritten as
\begin{equation}
\label{tail} \prod_{k\geqslant
N+1}^\infty\left(1-\frac{n^2+O\left(\frac{1}{n^2}\right)}{k^2}+O\left(\frac{1}{k^4}\right)\right)
\end{equation}
So we see that if $n$ is close to $N$, then the terms
$\frac{n^2+O\left(\frac{1}{n^2}\right)}{k^2}$ are close to one,
and consequently the first factors in (\ref{tail}) are small. This
implies that the product in (\ref{alpha_est}) is not close to one.
This simple observation yields that we can guarantee the good
estimate in (\ref{alpha_est}) only if $N$ is much greater than
$n$, so that the product in (\ref{tail}) is close to one.

We need to find out how many eigenvalues (we call their number by
$N$) we need to recover in order to $n$ less than $N$ approximate
normalizing coefficients be recovered with a good accuracy. In
other words, we need to find out the relationship between $N$ and
$n$ such that the product in (\ref{alpha_est}) or what is
equivalent (\ref{tail}) are close to one. Let us notice that if
$0<x<\theta<1$ then $|\ln{(1-x)}|<
\frac{|\ln{(1-\theta)}|}{\theta}x$. Using this observation, we can
estimate for $\frac{n^2}{(N+1)^2}\leqslant\theta$:
\begin{eqnarray}
\left|\ln\prod_{k\geqslant
N+1}^\infty\left(1-\frac{n^2}{k^2}\right)\right|\leqslant
\sum_{k\geqslant N+1}^\infty
\left|\ln\left(1-\frac{n^2}{k^2}\right)\right|\notag\\ \leqslant
\sum_{k\geqslant N+1}^\infty
\frac{|\ln{(1-\theta)}|}{\theta}\frac{n^2}{k^2}=n^2
\frac{|\ln{(1-\theta)}|}{\theta}\left(\frac{\pi^2}{6}-\sum_{k=1}^N\frac{1}{k^2}\right)\label{sum_est}
\end{eqnarray}
If we choose $N$ and $n$ such that right hand side of
(\ref{sum_est}) is less than some $\varepsilon>0$, then
\begin{equation}
e^{-\varepsilon}<\prod_{k\geqslant
N+1}^\infty\left(1-\frac{n^2}{k^2}\right)<1.
\end{equation}
Consequently, for such $N$ and $n$ we have (see
(\ref{alpha_est})):
\begin{equation}
\label{alpha_est1}
|\alpha_n-\widetilde\alpha_{n,N}|\leqslant|A_n||\widetilde
B_{n,N}|\left|1-e^{-\varepsilon}\right|.
\end{equation}
Notice that since $A_n=1+o(1)$ and $\alpha_n=\frac{1}{\sqrt{2}\pi
n}+o(1)$ as $n\to\infty$, we have that $|A_n||\widetilde B_{n,N}|$ is
bounded by some positive $C<4$. So, taking $\varepsilon$ small
enough, and picking $N$ and $n$ such that the right hand side of
(\ref{sum_est}) be less that $\varepsilon$, we arrive at
(\ref{alpha_est1}). Using formula
\begin{equation*}
\frac{\pi^2}{6}=\sum_{k=1}^N\frac{1}{N}-\frac{1}{2N^2}+O\left(\frac{1}{N^3}\right)
\end{equation*}
from \cite{TB} we summarize all our observations in the lemma:
\begin{lemma}
\label{Lemma_alpha_est} Let $0<\varepsilon<1$. If $n$ and $N$ satisfy
\begin{equation}\label{N_est}
\frac{n^2}{N}\leqslant\frac{\varepsilon}{2\ln{2}},
\end{equation}
then there exists an absolute constant $C>0$ such that
\begin{equation}
\label{alpha_est_1} |\alpha_k-\widetilde\alpha_{k,N}|\leqslant C
\left|1-e^{-\varepsilon}\right|, \forall k=1,\ldots,n
\end{equation}
\end{lemma}
Let us recall the following important
representations (see \cite{AM}) for the response function and for
the kernel $c^\tau$ (in our case $\tau=1$):
\begin{lemma}\label{lemrepresentation}
Assume that $q\in L^1(0,1)$. Then the following representation for the
response function $r$,
\begin{equation}
\label{Resp_mes_conn} r(t)=\int_{-\infty}^\infty
\frac{\sin{\sqrt{\lambda}t}}{\sqrt{\lambda}}\,d\sigma(\lambda),\,
\end{equation}
holds for almost all $t \in (0,2\tau)$.

The kernel $c^\tau(s,t)$ admits the following representation:
\begin{equation}
c^\tau(s,t)=\int_{-\infty}^\infty
\frac{\sin{\sqrt{\lambda}(\tau-t)}\sin{\sqrt{\lambda}(\tau-s)}}{\lambda}\,d\sigma(\lambda),\quad
s,t\in [0,\tau],
\end{equation}
where the integral in the right-hand side of (\ref{C_T_eqiv})
converges uniformly on $[0,\tau]\times [0,\tau]$.

\begin{equation}
\label{c_t} c^\tau(t,s)=p(2\tau-t-s)-p(t-s)
\end{equation}
and $p(t)$ is defined by
\begin{equation}
p(t):=\frac{1}{2}\int_0^{|t|} r\left(s\right)\,ds.
\end{equation}
\end{lemma}
The following useful formula follows directly from representations
(\ref{Resp_mes_conn}), (\ref{C_T_eqiv}):
\begin{equation}
\label{C_T_Diagonal}
c^\tau(t,t)=\frac{1}{2}\int_0^{2(\tau-t)}r(\tau)\,d\tau
\end{equation}
If the exact values of the first
$n$ eigenvalues $\lambda_k$ and normalizing factors $\alpha_k^2$,
$k=1,\ldots,n$, were known, then one could construct the
``restricted" response functions and kernels defined by
$$
r_n(t)=\sum_{k=1}^n
\left[\frac{\sin{\sqrt{\lambda_k}t}}{\sqrt{\lambda_k}}\frac{\operatorname{sign}{\lambda_k}}{\alpha_k^2}-
\frac{\sin{\sqrt{\lambda^0_k}t}}{\sqrt{\lambda^0_k}}\frac{1}{(\alpha^0_k)^2}\right],
$$
\begin{multline*}
\hspace{1cm}    c_n^\tau(t,s)=\sum_{k=1}^n\frac{\sin{\sqrt{\lambda_k}(T-t)}\sin{\sqrt{\lambda_k}(T-s)}}{\lambda_k}\frac{\operatorname{sign}{\lambda_k}}{\alpha_k^2}\\
-\frac{\sin{\sqrt{\lambda^0_k}(T-t)}\sin{\sqrt{\lambda^0_k}(T-s)}}{\lambda^0_k}\frac{1}{(\alpha^0_k)^2}.
\hspace{1cm}
\end{multline*}
then every $r_n\in C^\infty(0,1)$ and Lemma \ref{lemrepresentation} yields
\begin{eqnarray}
r_n(t)\to r(t),\quad {\text{for almost all $t\in (0,2)$}},\\
c_n^\tau(t,s)\to c^\tau(t,s),\quad {\text{uniformly on $(0,\tau)^2$}},\label{Kernel_conv}\\
c_n^\tau(t,t)\to c^\tau(t,t),\quad {\text{uniformly on the
diagonal}}.\label{Kernel_diag_conv}
\end{eqnarray}
As we only have access to approximate values of the normalizing factors in practice, we can only compute the approximate restricted kernel by
\begin{multline}\label{approx_restr_ker}
\hspace{1cm}\widetilde c_{n,N}^\tau(t,s)=\sum_{k=1}^n\frac{\sin{\sqrt{\lambda_k}(T-t)}\sin{\sqrt{\lambda_k}(T-s)}}{\lambda_k}
\frac{\operatorname{sign}{\lambda_k}}{\widetilde\alpha_{k,N}^2}\\
-\frac{\sin{\sqrt{\lambda^0_k}(T-t)}\sin{\sqrt{\lambda^0_k}(T-s)}}{\lambda^0_k}\frac{1}{(\alpha^0_k)^2}.
\hspace{1cm}
\end{multline}
Then we can estimate the difference
\begin{equation}
\label{C_n_est} \|c_n^\tau-\widetilde
c_{n,N}^\tau\|_\infty\leqslant\sum_{k=1}^n\frac{|\widetilde\alpha_{k,N}^2-\alpha_k^2|}{\lambda_k\widetilde\alpha_{k,N}^2\alpha_k^2}=
\sum_{k=1}^n\frac{|\widetilde\alpha_k+\alpha_k||\widetilde\alpha_k-\alpha_k|}{\lambda_k\widetilde\alpha_{k,N}^2\alpha_k^2}
\end{equation}
Using (\ref{alpha_est_1}) and asymptotical expansions for the
eigenvalues and norming coefficients (\ref{eigen_exp})
(\ref{norm_coeff_exp}), we deduce from (\ref{C_n_est}) that (below, $C$ denotes an absolute constant that might change from line to line):
\begin{equation*}
\|c_n^\tau-\widetilde c_{n,N}^\tau\|_\infty\leqslant C \sum_{k=1}^n\frac{|\widetilde\alpha_{k,N}^2-\alpha_k^2|}{\lambda_k\widetilde\alpha_{k,N}^2\alpha_k^2}\leqslant
C\sum_{k=1}^n k|\widetilde\alpha_{k,N}-\alpha_k|.
\end{equation*}
Having remembered estimate (\ref{alpha_est_1}), we finally get
\begin{equation}
\label{C_n_est_1} \|c_n^\tau-\widetilde c_{n,N}^\tau\|_\infty\leqslant
C\frac{n(n+1)}{2} \left|1-e^{-\varepsilon}\right|.
\end{equation}
Let us fix some $\delta>0$ and choose $n\in \mathbb{N}$ such that
$\|c_n^\tau-c^\tau\|_\infty\leqslant \dfrac{\delta}{2}$ and consider the
difference
\begin{equation}
\|\widetilde c_{n,N}^\tau-c^\tau\|_\infty\leqslant \|\widetilde
c_{n,N}^\tau-c_n^\tau\|_\infty+\|c_n^\tau-c^\tau\|_\infty
\end{equation}
Using (\ref{C_n_est_1}) we obtain the existence of a constant $C^*>0$ such that
\begin{equation}
\label{E_est} \|\widetilde c_{n,N}^\tau-c^\tau\|_\infty\leqslant C^*
n^2\left|1-e^{-\varepsilon}\right|+\frac{\delta}{2}.
\end{equation}
Then by choosing $\varepsilon$ in (\ref{E_est}) we achieve
\begin{equation}
\|\widetilde c_{n,N}^\tau-c^\tau\|_\infty\leqslant \delta.
\end{equation}
We summarize the above observations in the following Proposition, which details the main of our approximation procedure.
\begin{proposition}\label{Kernel_close}
Let $\delta>0$ be fixed. Let $n$ be chosen that (this is possible thanks to \eqref{Kernel_conv})
$$
\|c_n^\tau-c^\tau\|_\infty\leqslant \dfrac{\delta}{2}.
$$
Next, take $\varepsilon>0$ such that
$$
n^2\left|1-e^{-\varepsilon}\right|\leqslant\dfrac{\delta}{2},
$$
Finally, choose $N$ such that
$$
\frac{n^2}{N}\leqslant\dfrac{\varepsilon}{2\ln{2}}.
$$
Then, there exists an absolute constant $C>0$ such that the following estimate holds true:
$$
\|\widetilde c_{n,N}^\tau-c^\tau\|_\infty\leqslant C\delta,
$$
where the approximate restricted kernel $\widetilde c_{n,N}^\tau$ and the approximate normalizing
coefficients $\widetilde \alpha_{k,N}$, $k=1,\ldots,n$ are respectively defined by
(\ref{approx_restr_ker}) and (\ref{approx_coeff}).

In particular, $\widetilde c_{n,N}^\tau$ converges uniformly to $c^\tau$ when $n$ tends to infinity and $N$ is chosen as above.
%
%
\end{proposition}

Along with the equation (\ref{G_LN}) we consider the equation for
the approximate restricted kernel $\widetilde c_{n,N}^\tau$:
\begin{equation}
\label{G_LN_restr} \widetilde V_{n,N}(y,t)+\widetilde
c^\tau_{n,N}(y,t)+\int_y^\tau \widetilde c^\tau_{n,N}(t,s)\widetilde
V_{n,N}(y,s)\,ds=0,\quad 0<y<t<\tau.
\end{equation}
Picking $\delta>0$ we can use Proposition \ref{Kernel_close} to find $n$
and $N$ such that $\|\widetilde
c_{n,N}^\tau-c^\tau\|_{\infty}\leqslant\delta$. From now on, we always assume that $n$ and $N$ are chosen according to Proposition \ref{Kernel_close}. Note that in particular, we have $N\to +\infty$ as $n\to +\infty$.

We introduce the operator
$\widetilde \Cscr_\tau^{n,N}$ which are given by an integral part of
(\ref{r-c2}) with the kernel $c^\tau$ substituted by $\widetilde
c^\tau_{n,N}$. The closeness of $c^\tau$ and $\widetilde c^\tau_{n,N}$
implies the operator $I+\widetilde C_\tau^{n,N}$ to be invertible
along with $I+C_\tau$. The latter in turn implies the existence of
the potential $\widetilde q_{n,N}$ that produces the response function
\begin{equation}
\widetilde r_{n,N}(t)=\sum_{k=1}^n
\left[\frac{\sin{\sqrt{\lambda_k}t}}{\sqrt{\lambda_k}}\frac{\operatorname{sign}{\lambda_k}}{\widetilde\alpha_{k,N}^2}-
\frac{\sin{\sqrt{\lambda^0_k}t}}{\sqrt{\lambda^0_k}}\frac{1}{(\alpha^0_k)^2}\right],
\end{equation}
and the unique solvability of (\ref{G_LN_restr}) (see \cite{AM},
\cite{BI}, \cite{ABI}).


We denote $M:=\|(I+C_\tau)^{-1}\|$. The invertibility of
$I+C_\tau$ and $I+\widetilde C_\tau^{n,N}$ implies the norms of the
solutions $\|V(y,\cdot)\|_{L^2}$ and $\|\widetilde
V_{n,N}(y,\cdot)\|_{L^2}$ to be bounded.

Let us write down the difference (\ref{G_LN}) and
(\ref{G_LN_restr}) in the form
\begin{eqnarray}
\label{difference}
\left[V(y,t)-\widetilde V_{n,N}(y,t)\right]+\int_y^\tau c^\tau(t,s)\left[V(y,s)-\widetilde V_{n,N}(y,s)\right]\,ds\\
=\widetilde
c_{n,N}^\tau(y,t)-c^\tau(y,t)+\int_y^\tau\left[c^\tau(t,s)-\widetilde
c_{n,N}^\tau(t,s)\right]\widetilde V_{n,N}(y,s)\,ds,\quad
0<y<t<\tau.\notag
\end{eqnarray}
The above equality and the invertibility of $I+C^\tau$ implies the
estimate:
\begin{equation}
\label{L^2-norm} \|V(y,\cdot)-\widetilde
V_{n,N}(y,\cdot)\|_{L^2(y,\tau)}\leqslant M\left(1+\max_{0\leqslant
y\leqslant \tau}\|\widetilde
V_{n,N}(y,\cdot)\|_{L^2}\right)\|c^\tau-\widetilde c_{n,N}^\tau\|_\infty.
\end{equation}
To estimate the $L^\infty$ norm, we write down (\ref{difference}) in the
form
\begin{eqnarray}
\label{On_diagonal}
\left[V(y,t)-\widetilde V_{n,N}(y,t)\right]=\int_y^\tau c^\tau(t,s)\left[\widetilde V_{n,N}(y,s)-V(y,s)\right]\,ds\\
+\widetilde
c_{n,N}^\tau(y,t)-c^\tau(y,t)+\int_y^\tau\left[c^\tau(t,s)-\widetilde
c_{n,N}^\tau(t,s)\right]V_{n,N}(y,s)\,ds,\quad 0<y<t<\tau.\notag
\end{eqnarray}
The latter leads to the following inequality:
\begin{eqnarray}
\label{Inq_C}
\|V(y,t)-\widetilde V_{n,N}(y,t)\|_\infty\leqslant\|c^\tau\|_\infty\max_{0\leqslant y\leqslant\tau}\int_y^\tau\left|\widetilde V_{n,N}(y,s)-V(y,s)\right|\,ds\\
+\|\widetilde c_{n,N}^\tau-c^\tau\|_\infty+\|c^\tau-\widetilde
c_{n,N}^\tau\|_\infty\max_{0\leqslant y\leqslant\tau}\int_y^\tau|\widetilde
V_{n,N}(y,s)|\,ds.\notag
\end{eqnarray}
Using (\ref{L^2-norm}) and (\ref{Inq_C}) we finally get
\begin{equation}
\label{Inq_C_1} \|V-\widetilde V_{n,N}\|_\infty\leqslant
\|c^\tau-\widetilde c_{n,N}^\tau\|_\infty \left(M \|c^\tau\|_\infty+1\right)
\left(1+\max_{0\leqslant y\leqslant \tau}\|\widetilde
V_{n,N}(y,\cdot)\|_{L^2}\right) \notag
\end{equation}
The last equality in particular implies the uniform convergence of
$\widetilde V_{n,N}(y,y)$ to $V(y,y)$. Having remembered
(\ref{Rec_potenc_GL}), we conclude that the potentials $q_{n,N}$ and $q$ corresponding respectively to $\widetilde V_{n,N}$ and $V$ satisfy
\begin{equation}
\label{pot_conv} \int_0^{t}\widetilde
q_{n,N}(s)\,ds\rightrightarrows_{n\to\infty}
\int_0^{t}q(s)\,ds,\quad\text{uniformly in $t$.}
\end{equation}
The latter in turn, implies that
\begin{equation}
\widetilde q_{n,N}\longrightarrow q,\quad \text{in $H^{-1}(0,1)$}.
\end{equation}
In fact, (\ref{pot_conv}) yields more than that: we have
\begin{equation}
\label{pot_conv_1}
\frac{1}{\varepsilon}\int_t^{t+\varepsilon}\widetilde
q_{n,N}(s)\,ds\rightrightarrows_{n\to\infty}
\frac{1}{\varepsilon}\int_t^{t+\varepsilon}q(s)\,ds,\quad\text{uniformly
in $t$, $\varepsilon$.}
\end{equation}
We remark that (\ref{pot_conv_1}) is still not enough to guarantee
the convergence $\widetilde q_{n,N}$ to $q$ almost everywhere on
$(0,1)$.

On the other hand let us restrict (\ref{G_LN}) to the diagonal
$y=t$:
\begin{equation}
\label{G_LN_diagg}
V(y,y)+c^T(y,y)+\int_y^Tc^T(y,s)V(y,s)\,ds=0,\quad 0<y<1,
\end{equation}
and recalling (\ref{Rec_potenc_GL}), (\ref{C_T_Diagonal}), we see
that the best possible result one can expect is a convergence
$\widetilde q_{n,N}$ to $q$ almost everywhere on $(0,1)$.

\begin{remark}
The stability of the scheme crucially depends on the type of the
convergence $r_n\to r$. For now we know (see \cite{AM}) that the
convergence is pointwise almost everywhere on the interval. The
significant progress in the proving of the stability could be
achieved by the improving of this result.
\end{remark}


\bibliographystyle{siam}
\bibliography{InverseSchrodinger}

\end{document}